  \documentclass[twocolumn]{svjour3}          

  \usepackage{times}
  \usepackage{subfig}
  \usepackage[dvipdfmx]{color}
  \usepackage{comment}
  \usepackage{color}
  \usepackage[pdftex]{graphicx}
  \usepackage{url}
  \usepackage{amsmath,amssymb,amsfonts}
  \usepackage{algorithmic}
  \usepackage{algorithm}

  \def\Vec#1{\mbox{\boldmath $#1$}}
  
  \newcommand{\onote}[1]{{\color{blue} #1 \color{black}}}       
  \newcommand{\del}[1]{}                                        
  \newcommand{\jtext}[1]{{\color{blue} #1 \color{black}}}       
  
  \newcommand{\jtextd}[1]{}
  \newcommand{\ocut}[1]{}

  \def\Vec#1{\mbox{\boldmath $#1$}}

  \graphicspath{{./Figs/}}
  \RequirePackage{fix-cm}
  \smartqed  
  \usepackage{graphicx}
\begin{document}
  
  \title{%
  Adjust-free adversarial example 
  generation in speech recognition using 
  evolutionary multi-objective optimization
  under black-box condition
  }
  
  
  \author{
    Shoma Ishida \and
    Satoshi Ono.
  }
  
  
  \institute{Shoma Ishida \at
                Department of Information Science
                and Biomedical Engineering,
                Graduate School of Science
                and Engineering,
                Kagoshima University, Japan\\
                \email{sc115002@ibe.kagoshima-u.ac.jp}           
             \and
             Satoshi Ono \at
                Department of Information Science
                and Biomedical Engineering,
                Graduate School of Science
                and Engineering,
                Kagoshima University, Japan\\
                \email{ono@ibe.kagoshima-u.ac.jp}
  }
  
  \date{Received: date / Accepted: date}

  \maketitle
  
  \begin{abstract}
      \jtextd{
      本論文では，ブラックボックス条件下で実行する進化型多目的最適化（EMO）ベースの頑健な敵対的音声事例の生成手法を提案する．従来手法では，対象モデルの内部情報を利用するホワイトボックス条件下での敵対的事例の生成に焦点あてている．
      提案された手法は，EMOの母集団ベースの探索と目的関数を複数設定できる特性により，
      頑健な敵対的事例の生成する．実験結果は，時間軸方向に頑健な敵対的事例の生成に成功したことから，提案された手法の可能性を示した．
      }
      
      This paper proposes a black-box adversarial attack method to
      automatic speech recognition systems. 
      %
      Some studies have attempted to attack neural networks for speech recognition;
      however, these methods 
      did not consider the robustness of generated adversarial
      examples against timing lag with a target speech.
      The proposed method in this paper
      adopts 
      Evolutionary Multi-objective Optimization (EMO) 
      that allows it generating robust adversarial examples under black-box scenario.
      %
      Experimental results showed that the proposed method successfully generated
      adjust-free adversarial examples, which are sufficiently robust
      against timing lag so that
      an attacker does not need to take the timing of playing it against the target speech.
      %
      %
     
  \end{abstract}

  \thispagestyle{empty}
  
  
  \section{INTRODUCTION}
  
  \jtextd{近年，深層ニューラルネットワーク（深層 NN）を用いた自動音声認識（ASR）システムは，スマートフォンのパーソナルアシスタント（Amazon Alexa，Apple Siri）からデバイス（Amazon Echo，Google Home）まで，多くの製品で広く使用されている．一方で深層 NNは，サンプルに小さな摂動を意図的に加えることによって生成された敵対的事例\cite{goodfellow2014explaining}に対して脆弱であることが知られている。ASRシステムは気密性の高いユーザーデータを必要とする複雑なタスクも実行することができるようになったことから，高いセキュリティを備えることが重要である．そのため，ASRシステムの堅牢性を評価および改善する目的から，敵対的事例を生成する研究が行われている．}
  
  In recent years, Automatic Speech Recognition (ASR) systems are widely used in many products such as personal assistants of smartphones,
voice command technologies in cars, and so on. On the other hand, deep learning methods are known to be vulnerable to adversarial examples, small perturbations added to target samples~\cite{goodfellow2014explaining}. It is essential that ASR systems have high security because ASR systems perform various tasks which may require user personal data. Therefore, studies have been conducted to generate adversarial examples to improve the robustness of ASR systems.
  
  \jtextd{ASRに対するAEの生成は，ホワイトボックス条件，すなわち．．．が多
    かった． 近年は，ブラックボックス条件も増えている． 実環境においてよ
    り深刻となる脅威についての研究が行われるようになってきた．さらに，頑
    健なAEは．．．}
  
  Early studies of adversarial attack to ASR
  systems~\cite{yakura2018robust,carlini2018audio}
  were based on white-box conditions, where
  internal information of a target Deep Neural Network (DNN) classifier such
  as gradients of a loss function is available;
  however, most consumer ASR systems and services prohibit the access to
  their internal structure and parameters.
  Therefore, the white-box conditions are not realistic in terms of
  analyzing the safety of the consumer systems.
  Recently, a few studies attempted to attack the ASR systems under
  black-box condition where classification result (class labels) and its
  confidence are available but internal information is not~\cite{alzantot2018did,taori2019targeted}.
  %
  The black-box attack can be applicable consumer ASR 
  services
  and it is
  expected to find their vulnerabilities.
  %
  On the other hand, to discover more serious vulnerabilities in the
  real world, it is indispensable to design robust perturbations against
  environmental changes, time gap between the target speech and
  perturbation, and so on.

  \jtextd{そこで本研究では，ブラックボックス条件下で頑健な敵対的音声摂動の生成方式を提案する．提案方式では，頑健な敵対的事例の生成問題を多目的最適化問題として定式化し，進化型多目的最適化アルゴリズムにより解くことで，時間軸方向のズレに対してタイミングの調整が不要でかつ，頑健な敵対的音声摂動の生成を可能にする．
  このような例を「アジャストフリー摂動」と呼びます．
  音声コマンド分類モデル\cite{sainath2015convolutional}を対象とした評価実験を行い，提案方式と従来方式により生成された敵対的事例とを，時間軸方向のズレに対する頑健性の観点で比較し，提案方式の有効性を確認した．}
  
  Therefore, this paper proposes a method for generating adversarial
  examples to ASR systems that are robust
  against time difference.
  In the actual environment, it is difficult for an attacker to play
  the perturbation noise accurately in time with the target speech,
  and a time difference significantly reduces the effect of the
  attack.
  In other words, robust adversarial perturbations, which cause
  misrecognition even when the timing is slightly off, are more
  threatening.
  Creating such robust adversarial noises makes it easier for the
  attacker to determine when to start the noise.
  Moreover, an adversarial example that leads the classifier
  to misrecognize it even with the arbitrary length of the time difference
  makes it unnecessary for the attacker to pull the trigger; the
  attacker only needs to repeat the adversarial noise sound
  like an environmental sound. 
  We term such a highly threatening example an {\it adjust-free}
  adversarial example.
  In the proposed method, adjust-free
  adversarial example design is formulated
  as a multi-objective optimization problem, and an evolutionary
  multi-objective optimization algorithm solves the problem. 
%
  %
  Experiments using speech commands classification
  model~\cite{sainath2015convolutional} showed 
  that
  the proposed method
  successfully generated robust adversarial examples 
  against the time difference and also adjust-free adversarial examples.

  \section{RELATED WORK}

  \jtextd{yakuraらは，残響と再生環境からのノイズを考慮し，物理的な世界でASRモデルに対して頑健な敵対的事例の生成方法を提案した\cite{yakura2018robust}．この手法では，物理世界での再生または記録によって引き起こされる変換プロセスを組み込むことによって，敵対的事例の生成を行う．しかし，この手法は学習器の内部情報（損失関数の勾配や重みパラメータ等）を利用するホワイトボックス条件を仮定しており，商用ASRシステムに適用することが難しい．
  
  一方で近年は，学習器の内部情報を利用できず，識別結果とその信頼度のみが
  参照可能となるブラックボックス条件下で敵対的事例を生成する方法が提案さ
  れている．ブラックボックス条件下では，目的関数の勾配を必要としない進化
  計算を利用することで敵対的摂動を生成する方式が提案されている
  \cite{su2019one}．
  
  敵対的摂動を生成する問題は本質的に，分類精度と摂動量のトレードオフ関係にあり，多目的最適化を行うことで敵対的摂動を生成する方式が提案されている\cite{suzuki2019adversarial}．一方，画像認識を対象とした敵対的事例の生成の分野では，角度、視点、照明条件に対して頑健な敵対的事例の研究が行われている\cite{athalye2017synthesizing}．音声認識に対する敵対的攻撃を行う際は，現実世界でノイズと音声の干渉時にズレが発生する状況が予想できるが，攻撃対象となる発話に対して摂動を加えるタイミングがずれることにより攻撃の効果が低下する．言い換えると，多少タイミングがずれた場合にも誤認識を引き起こす敵対的事例は，より脅威の度合いが高いといえる．このような事例を，本研究では頑健な敵対的事例と呼ぶ．従来の音声認識の脆弱性を検証する研究では時間軸方向のずれに頑健な敵対的事例の検討が行われていない．
  }
  
  Yakura has proposed a method for generating robust adversarial
  examples for ASR systems in physical environments, which considers the
  reverberation and noise from the regenerative
  environment~\cite{yakura2018robust}. This method generates adversarial
  examples by simulating a transformation process of playback or
  recording in the physical environments. However, it assumes
  the white box setting
  and therefore it is difficult to apply this method to
  commercial ASR systems.
  
  Recently, a few studies proposed adversarial attack methods
  under black-box setting where internal information of
  a target classifier cannot be available and only class labels and
  their confidence scores can be referred. For instance, Su et
  al. proposed a method to generate adversarial perturbations using
  evolutionary computation, which does not require gradients of
  an objective function~\cite{su2019one}.
  %
  Khare et al. proposed a method to generate adversarial perturbations
  using Evolutionary Multi-objective Optimization (EMO) that controls the
  balance between conflicting objectives such as the text dissimilarity
  and acoustic similarity ~\cite{khare2018adversarial}.

  In the field of image recognition, studies on adversarial examples
  robust to image translations caused by changes of viewpoints and
  lighting conditions have been
  conducted~\cite{athalye2017synthesizing}.
  %
  Suzuki et al. proposed a method of generating adversarial perturbations
  using
  EMO~\cite{suzuki2019adversarial},
  %
  which optimizes the expectation and
  variance of classification accuracy to design robust adversarial
  examples against image transformation.

  On the other hand, in the real environment of speech recognition, it
  is expected that a timing gap may occur between a target voice sound
  and a designed adversarial perturbation.
  Then, generating robust adversarial examples against the timing
  difference is necessary to find vulnerabilities that could be a
  threat in the real situation.
  Although there are few studies that attempt to design different kinds
  of robust adversarial examples for speech
  recognition~\cite{qin2019imperceptible},
  further studies that design robust adversarial examples are necessary
  to enhance the safety of ASR systems based on DNN.

  %
  
  \jtextd{敵対的事例を生成する最も一般的なアプローチは，対象となる NNモデルの内部情報である損失関数の勾配を利用する，ホワイトボックス条件下で生成する方式である．
  音声の領域でも，ホワイトボックス条件下で反復最適化ベースの攻撃を用い，敵対的事例を生成する[carlini2018]．
  また，近年は物理的な世界でASRモデルに対して頑健な敵対的事例の生成が可能であることが示されている[yakura2018]．
  この手法では，物理世界での再生または記録によって引き起こされる変換プロセスを組み込むことによって，敵対的事例の生成を行う．
  一方で近年は，学習器の内部情報を利用できず，識別結果とその信頼度のみが参照可能となるブラックボックス条件下で敵対的事例を生成する方法が提案されている．
  こちらに関しても音声の領域で，遺伝的アルゴリズムを用いて敵対的事例の生成が行われている[alzantot2018]．
  また，進化型多目的最適化を～～～
  
  }

  \section{THE PROPOSED METHOD}
  \subsection{Key ideas}
  
  \jtextd{3.1 基本アイデア
  本研究では，ブラックボックス条件下で\del{多様な}AE を同時に生成する手法を提案する．提案する方式の基本アイデアを以下に示す．
  
  1. 多目的最適化としての敵対的事例設計問題の定式化：
  敵対的事例の生成には本質的に，分類精度と摂動量などのトレードオフ関係を持つ複数の目的関数が含まれる．
  したがって，これらの目的関数を単一の目的関数に統合せずに，多目的最適化問題としてモデル化し，パレート解集合の発見を試みることは妥当なアプローチであると考える．
  提案手法は，目的関数を統合するためのパラメータを必要しない点，および，微分不可能かつ非凸形状の目的関数を最適化できる点に利点がある．
  一度の最適化によりパレート解集合を発見することで，対象音声の特性に応じた，かつ，最も用途に合致した AE を選択できる．

  2. 進化型多目的最適化（EMO）アルゴリズムの適用：
  提案手法は，多目的最適化を実行するために EMO アルゴリズムを採用する．
  代替モデルを訓練するアプローチと比較すると，提案手法は代替モデルを訓練する必要はない点に利点がある．
  さらに，EMOは多点探索であるため，対象音声に適した敵対的事例を網羅的に生成できる．
  なお，提案手法は先行研究よりも優れた敵対的事例を生成するとは限らないが，提案手法を用いて様々な敵対的事例を見つけることで，対象となる深層NNモデルの特性をより深く知ることや未知の攻撃パターンを理解することが可能となる．
  特に，EMO を適用することで，提案手法では，微分不可能，多峰性の関数を含め，様々な種類の目的関数および制約条件を使用できる．
  提案手法は，分類精度の期待値に加えて分類精度の標準偏差を目的関数に加えることによって，時間軸方向に対して頑健な敵対的事例を生成することができる．
  
  ３．ブラックボックスアプローチ：
  EMOの利点の1つである母集団ベースの探索を採用すると、提案された手法はターゲットモデルの勾配情報を必要とせず，分類ラベルと信頼度を含む分類結果を用いたブラックボックス設定[3]、[5]、[6]で実行される．
  したがって，提案された方法はNN以外の独自のシステムおよびモデルに適用可能である．
  }
  
  This paper proposes a black-box adversarial attack method to
  speech recognition using DNN.
  Followings are the key ideas:
  
  \begin{enumerate}
  
  \item {\bf Formulating as a Multi-Objective Optimization (MOO)
    problem:} Problems of generating adversarial examples against DNN
    essentially involves multiple objective functions
    that have a trade-off relationship, e.g., perturbation amount and
    classification performance --- it is difficult to design
    imperceptive perturbations that mislead DNN while ones involving
    large perturbation can be easily created. Generally, such competing
    criteria are combined with a weighted linear sum; however, it is
    natural to solve the problem without integrating them into one
    single objective function. The proposed method formulates the
    adversarial example generation problem as an 
    MOO problem. Therefore, the proposed method does not need a
    parameter to integrate objective functions.
  
  \item {\bf Applying EMO
    algorithm: } Because the proposed method solves the MOO problem
    using an EMO algorithm, objective functions
    can be
    flexibly designed; non-differentiable, non-convex and noisy
    functions are available.
   
  \item {\bf Finding more serious vulnerabilities via robust
    optimization:} 
    In the case of speech recognition, finding robust adversarial
    examples against timing lag is important because
    it is quite difficult
    to add perturbation to a target speech with accurate timing in
    actual environments.
    Therefore, taking advantages of EMO,
    the proposed method performs robust
    optimization
    by
    simultaneously minimizing the expected value of classification
    accuracy and its variance~\cite{suzuki2019adversarial}. 
    In particular, this study 
    aims to generate an adjust-free
    adversarial example, which fools the classifier even with the
    arbitrary length of time difference against a target speech.
  %

  \end{enumerate}
  
  \begin{figure}[t]
    \centering
    \includegraphics[width=0.5\textwidth,clip]{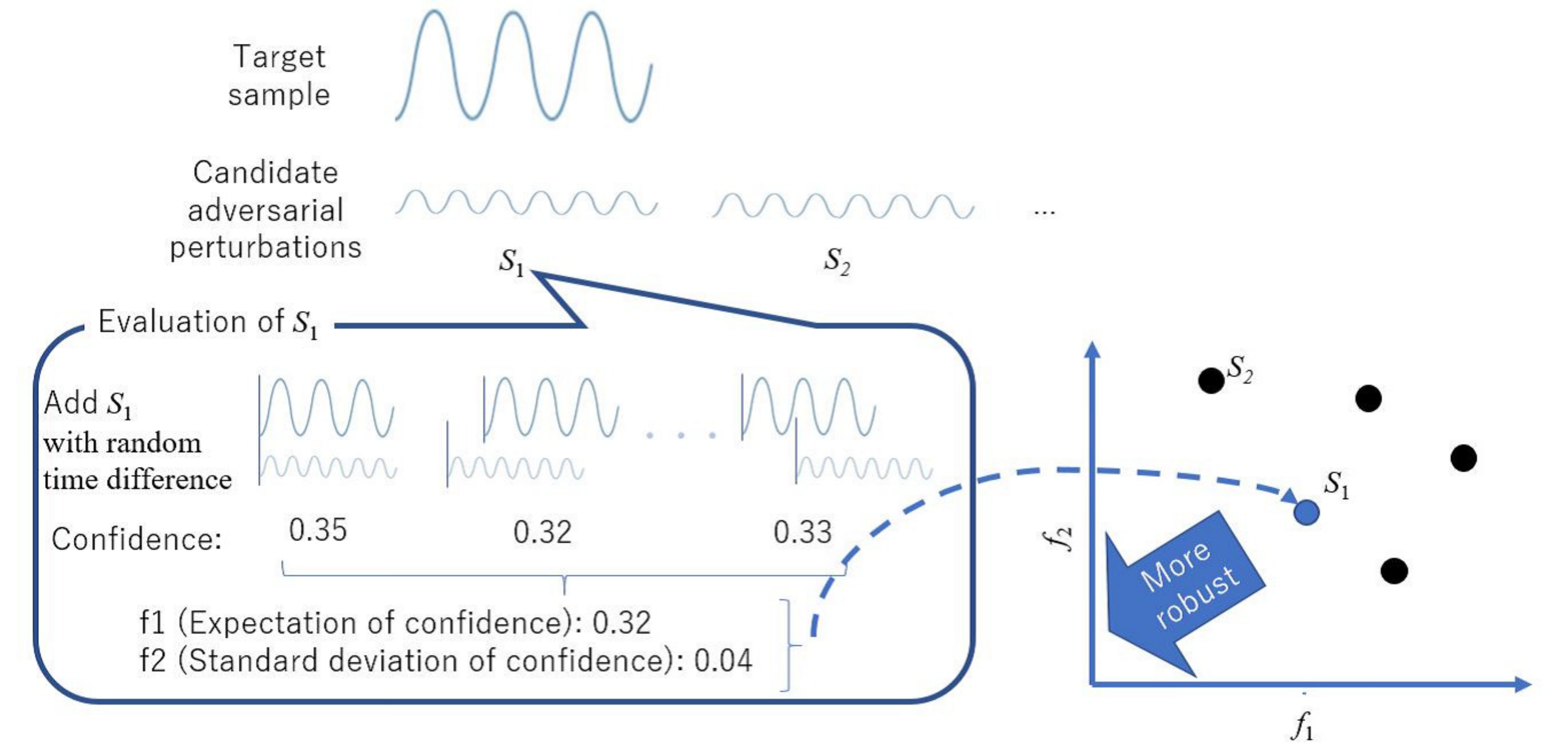}
    \caption{Robust adversarial example generation by multi-objective optimization}
    \label{fig:scheme}
~\\ ~\\  
    \begin{algorithm}[H]
      \caption{Adversarial example generation using MOEA/D}
      \label{alg:whole}
      \begin{algorithmic}[1]
        \STATE Generate initial population
        $\Vec{x}_1, \ldots, \Vec{x}_{N_{pop}}$
  %
        \STATE Select $N_f$ best individuals and subproblems
        $\mathcal{I}$.
  %
        \STATE Select mating and update range.
        \STATE Apply crossover and mutation operator.
        \STATE Evaluate offsprings.
        \STATE Update the reference point.
        \STATE Update the popoulation.
        \IF{stop condition is not satisfied}
        \STATE Go back to Step 2.
        \ENDIF
      \end{algorithmic}
    \end{algorithm}
    \vspace*{-10mm}
    \caption{Algorithm of the proposed method using MOEA/D.}
    \label{fig:whole}
  \end{figure}

  \subsection{Formulation}
  \jtextd{
  提案手法では，データセットの各オーディオサンプルについて、100個の個体が作成され、初期母集団が形成する．
  それらは，16,000のサンプリングレートでサンプリングした対象音声に対して，ランダムな均一摂動を加えることで初期化される．
  \onote{【1秒の音声を対象とする点や，個体数などは4章の実験条件で述べる．
      提案手法はそのような条件に限定されるものではなく，より一般的な手法
      であるため．】}
  
  ここで，解候補xは，以下のような変数～～から構成される．
  
  ～～設計変数の式～～
  
  ここで，ｔ　は音声における時間のサンプリング点を表す．
  
  }
  
  The proposed method designs an adversarial noise by optimizing perturbation amount
  to a sound wave amplitude of a target speech.
  For instance, if the a target speech length is one second and
  sampling rate is 16 kHz, the number of design variables, or the number
  of dimensions, gets into 16,000.
  %
  %
  
  \begin{figure*}[t]
    \centering
    \begin{tabular}[t]{@{}c@{}c@{}c@{}c@{}c@{}}
    \includegraphics[clip,width=32mm]{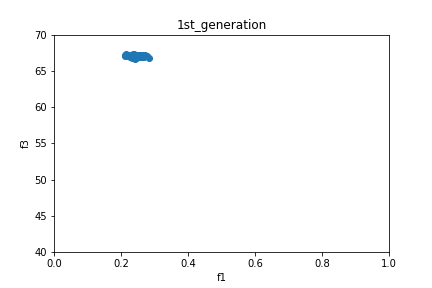} &
    \includegraphics[clip,width=32mm]{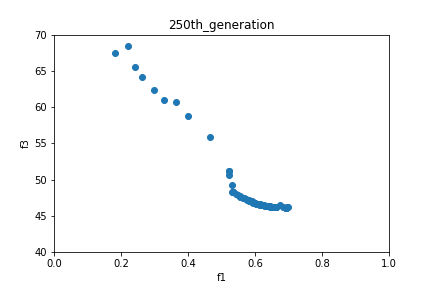} &
    \includegraphics[clip,width=32mm]{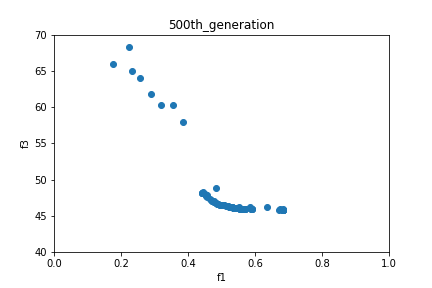} &
    \includegraphics[clip,width=32mm]{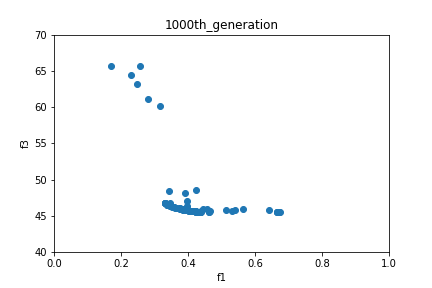} &
    \includegraphics[clip,width=32mm]{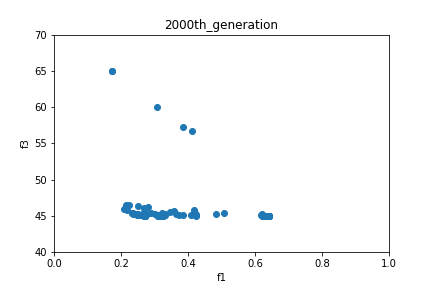} \\
    {\small (a) 1st generation} &
    {\small (b) 250th generation} &
    {\small (c) 500th generation} &
    {\small (d) 1,000th generation} &
    {\small (d) 2,000th generation} \\
  %
  %
    \end{tabular}
    \caption{Trasition on distributions of the non-dominated solutions
    for "down" sample.}
    \label{fig:transition}
  \end{figure*}
  
  ~\\
  \begin{figure}[t]
    \centering
    \begin{tabular}{@{}c@{~}c@{~}c@{}}
    \includegraphics[clip,width=40mm]{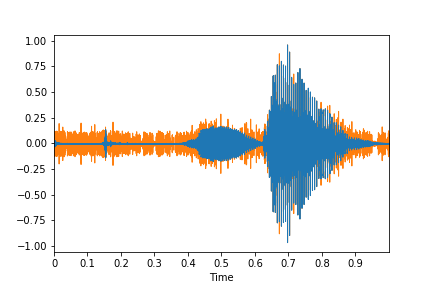} &
    \includegraphics[clip,width=40mm]{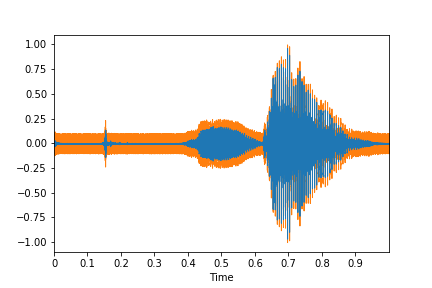} \\
    {\small (a) previous method} &
    {\small (b) proposed method } \\
    \end{tabular}
    \caption{Generated adversarial examples for class "down".}
    \label{fig:generated_example}
  \end{figure}
  
  \jtextd{
  提案手法は EMO を適用するため，目的関数や制約条件を柔軟に設定でき，かつ，複数の目的関数を同時に最適化することができる．
  従来の敵対的事例の生成では，対象音声が正しく認識される際の信頼度と，対象音声と生成した敵対的事例との音響的類似性の維持を 2 つの目的関数とする．
  これにより，対象音声における誤認識率と音響的類似性との関係性を明確化することができ，また，多様な敵対的事例を一度の最適化で生成することができる．
  本手法で取り組む頑健な敵対的事例の生成シナリオは、進化的計算の特性を活用した最適化の1つである[～]、[～]．
  従来研究では，ホワイトボックス設定[～]、[～]、[～]に基づいており、平均分類精度[～]、[～]のみを最小化していた．
  しかし，分類精度の平均値での最適化を行うと，特定の条件下で敵対的事例が正しく分類される可能性がある．
  そこで，本研究では，頑健な敵対的事例の生成を行うために，平均分類精度と音響的類似性に加え，標準偏差を第３の目的関数とすることで，このような誤分類の例外的な失敗を防ぎ、より頑健な敵対的事例を生成する．
  
  
  minimize f2 = RMSE(MFCC（I+x）- MFCC(I))
  
  minimize f3 = σ（P(C(I+x)=C(I))）
  
  ここで，C(·) は分類結果を表し，第 1 目的関数 f1 は，対象となる分類器が敵対的事例 I + x を正しいクラス C(I) に分類する信頼度の平均を示す．
  第 2 目的関数 f2 は，MFCC特徴 [～～]間の対象音声 I と生成された敵対的事例 I + x の平均二乗誤差平方根 RMSE(・)を用いて定量化する．
  MFCC特徴の抽出は、25ミリ秒のウィンドウサイズと10ミリ秒のストライドで実行する．
  第 3 目的関数 f3 は，～～～の標準偏差 σ（・）を示す．
  
  }
  
  The proposed method utilizes EMO-based approach, which allows
  designing objective functions and constraints flexibly and eliminating
  the need to combine objective functions into one function.
  To design adversarial examples robust against difference in timing
  with a target speech, the following three objective functions are designed:
  \begin{eqnarray}
  {\rm minimize}~ &&f_1(\Vec{x})  = 
              \mathbb{E}\left( 
               P( \mathcal{C} \left( \Vec{S} + \tau_i( \Vec{\rho} ) \right) 
                    = \mathcal{C}(\Vec{S}))
                \right)
  \nonumber \\
  {\rm minimize}~ &&f_2(\Vec{x}) = 
                     \sigma\left(
               P( \mathcal{C} \left( \Vec{S} + \tau_i( \Vec{\rho} ) \right) 
                    = \mathcal{C}(\Vec{S}))
             \right)
  \nonumber \\
  %
  {\rm minimize}~ &&f_3(\Vec{x}) =   || MFCC(\Vec{S} + \Vec{\rho}) - MFCC(\Vec{S}) ||_2
  \nonumber \\
  \label{eq:obj_func}
  \end{eqnarray}
  where 
  $\Vec{S}$ and $\Vec{\rho}$  denote a target speech and an adversarial perturbation,
  $\mathbb{E}(\cdot)$ and $\sigma(\cdot)$ are expected value and
  standard deviation of classification accuracy, and $\tau_i(\cdot)$
  denotes random timing lag within $\pm T_{max}$ second, i.e., the
  perturbation $\Vec{\rho}$ is played with a lag within $\pm T_{max}$
  second from target speech $\Vec{S}$.
  Previous work under white-box condition optimizes only the expected
  value of the confidence score~\cite{qin2019imperceptible}; 
  however, this might create AEs that
  could be correctly classified under a certain rare condition because
  such rare cases cannot be represented the averaged confidence.
  To avoid generating AEs that rarely cannot mislead classifiers, it is
  possible to use the worst (highest) confidence score among some tested
  cases as an objective function;
  however, such objective function is not smooth and cannot express
  slight improvement of solution candidates during optimization.
  Therefore, as shown in Fig.~\ref{fig:scheme}, we added the standard deviation of the classification
  accuracy as the second objective function~\cite{Ono2009_IJICIC}.

  
  \subsection{Process flow}
  \jtextd{
  本研究では
  式(\ref{eq:obj_func})のように定式化した多目的最適化問題に，
  MOEA/D\cite{zhang2007moea}を適用する．
  MOEA/Dは
  パレート解集合を発見する多目的最適化問題を，スカラー化関数を適用するこ
  とで単目的最適化問題の集合へと変換する．
  単目的化された各目的関数を定義する重みベクトル$\Vec{\lambda}^1, \ldots
  \Vec{\lambda}^{N_f}$に対して近傍を定義し，解候補の更新や選択を行う範囲
  を限定する．
  %
  \onote{【上記のような感じで，MOEA/Dの処理手順を簡単に説明しましょう．具体的なズレの量などは実験の章で述べる】}
  
  }

  %
  %
  
  
  This paper adopts Multi-Objective Evolutionary Algorithm based on
  Decomposition (MOEA/D)~\cite{zhang2007moea} to solve the
  MOO
  problem shown in eq.~(\ref{eq:obj_func}).
  Here, an original MOO problem is described as  follows:
  \begin{eqnarray}
   {\rm minimize} && \Vec{f}(\Vec{x}) = \left( f_1(\Vec{x}),f_2(\Vec{x}),f_3(\Vec{x})  \right)
    \nonumber \\
   {\rm subject~to} && \Vec{x} \in \Vec{\mathcal{F}}
  \end{eqnarray}
  %
  %
  In this paper, the above problem can be decomposed into many single objective optimization problems using the Tchevycheff method 
  as follows:
  \begin{eqnarray}
   {\rm minimize} && g(\Vec{x} | \Vec{\lambda}^j, z^*) 
                     = \max_{1 \leq i \leq N_f}  
                       \left\{ \lambda_i^j | f_i(\Vec{x}) - z_i^*  \right\}
    \nonumber \\
   {\rm subject~to} && \Vec{x} \in \Vec{\mathcal{F}}
   \end{eqnarray}
  where $\Vec{\lambda}^j = (\lambda_1^j, \ldots, \lambda_{N_f}^j )$ are weight
  vectors ($\lambda_i^j \geq 0$
  ) and 
  $\sum_{i=1}^{N_f} \lambda_i^j = 1$, and 
  $\Vec{z}^*$ is a reference point calculated as follows:
  \begin{equation}
  z_i^* = \min\{ f_i(\Vec{x}) |  x \in \Vec{\mathcal{F}} \} 
  \label{eq:reference_point}
  \end{equation}
  %
  %
  By preparing $N_D$ weight vectors and optimizing $N_D$ scalar objective
  functions, MOEA/D finds various non-dominated solutions at one
  optimization.
 
  %
  Figure~\ref{fig:whole} shows the detailed algorithm of the proposed method based on MOEA/D.
  The initial solution candidates $\Vec{x}_1, \ldots, \Vec{x}_{N_{pop}}$
  are generated by sampling them at uniformly random
  from the entire search space.
  %
  Then, 
  $N_f$ ($=3$) best individuals are selected for $N_f$ objective functions
  respectively, 
  and 
  the indexes of
  the subproblems $\mathcal{I}$ are selected.
%
  To reproduce new solution candidates, 
  crossover operator in 
  Differential Evolution (DE)~\cite{storn1997differential} 
  and polynomial mutation
  are applied to each $i \in \mathcal{I}$.
%
  The generated new solution candidates are evaluated by applying a
  target speech recognition model and by calculating objective
  function values from the classification result $\mathcal{C}(\Vec{S}
  + \Vec{\rho})$ with confidence scores.
  Then, the reference point and the population are updated according
  to the amount of constraint violations and values of $g$.
%
  %
  %
  After
  iterating $N_G$ generations, the algorithm stops.
  %
  %

  
    
  
  \begin{figure*}[t]
    \centering
    \begin{tabular}[t]{@{}c@{}c@{}c@{}c@{}c@{}}
    \includegraphics[clip,width=35mm]{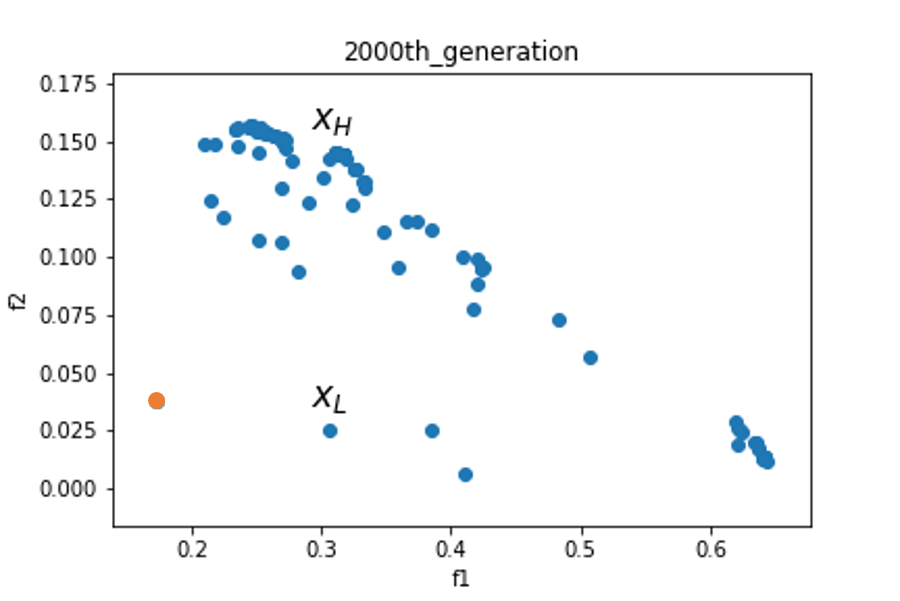} &
    \includegraphics[clip,width=35mm]{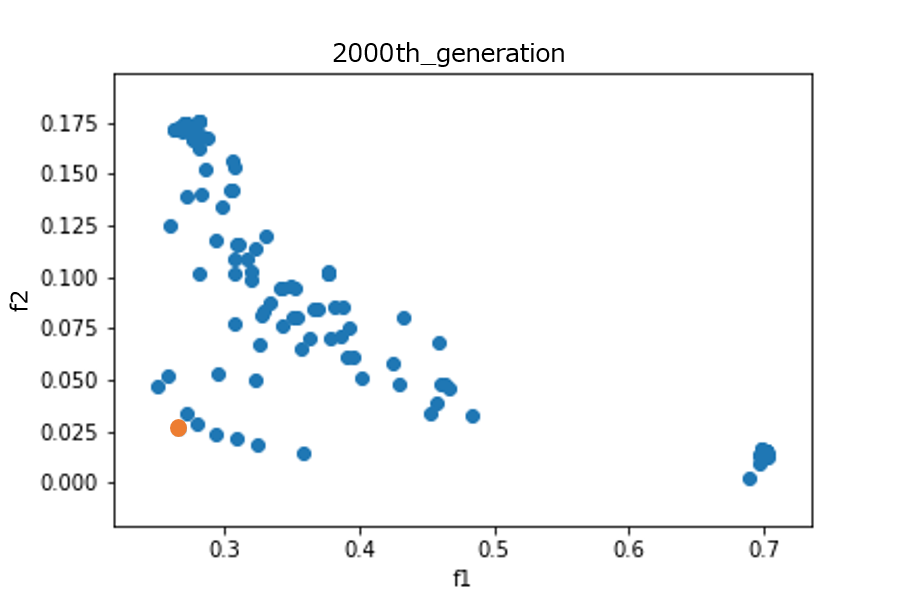} &
    \includegraphics[clip,width=35mm]{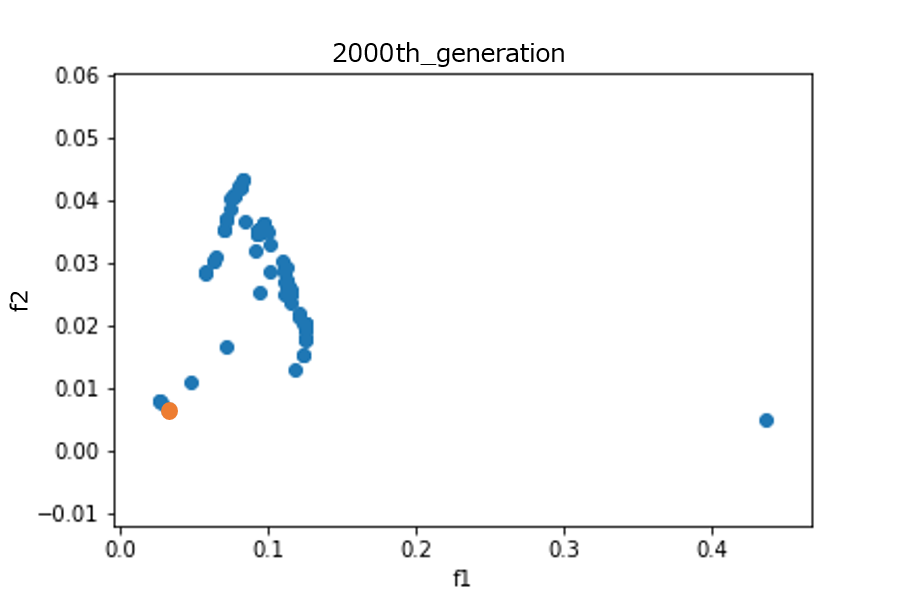} &
    \includegraphics[clip,width=35mm]{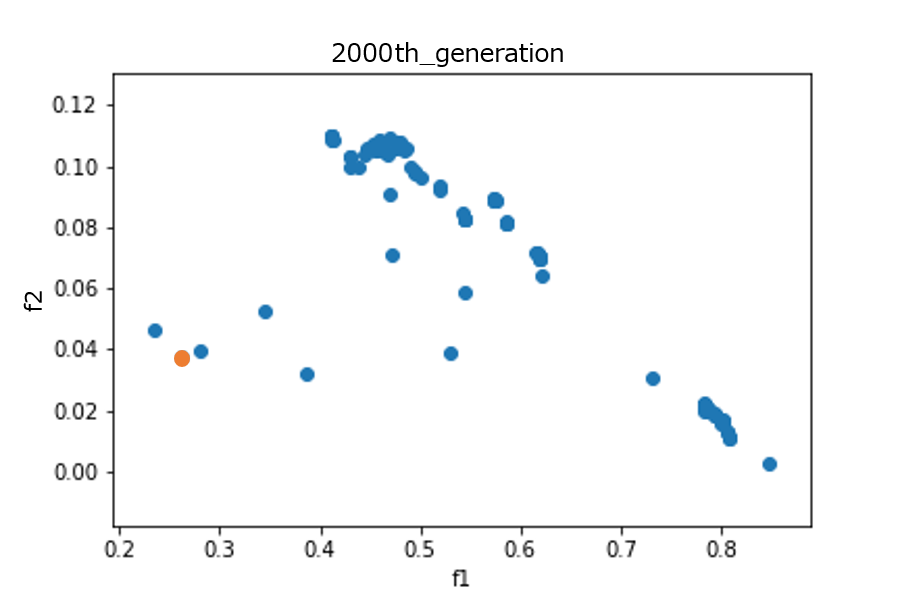} &
    \includegraphics[clip,width=35mm]{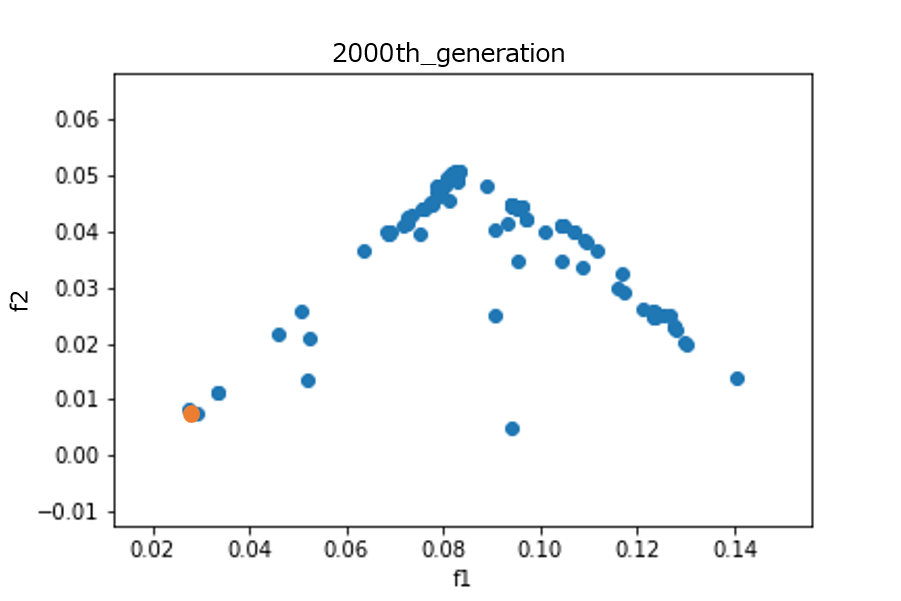} \\
    {\small (a) ``down'' } &
    {\small (b) ``no'' } &
    {\small (c) ``off'' } &
    {\small (d) ``on'' } &
    {\small (e) ``right'' } \\
    \includegraphics[clip,width=35mm]{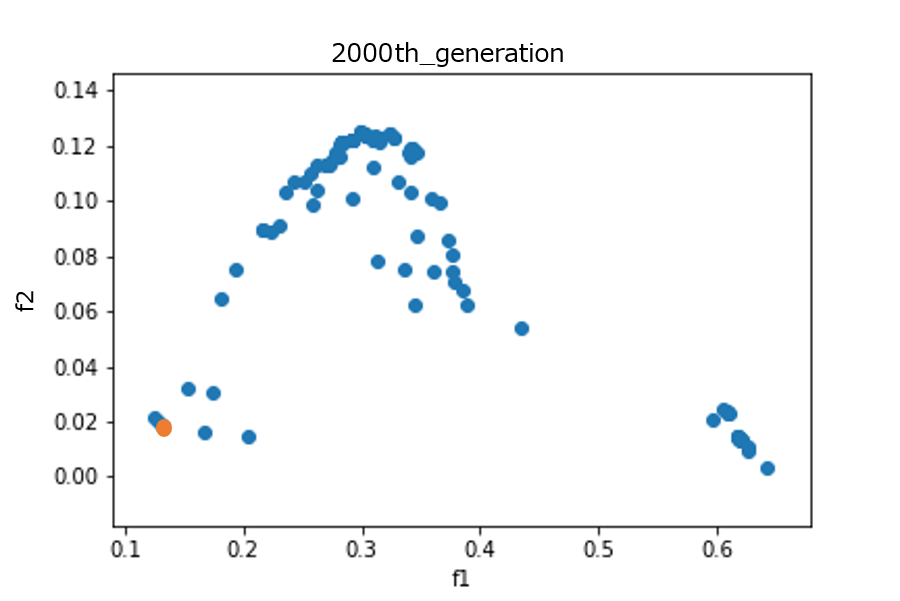} &
    \includegraphics[clip,width=35mm]{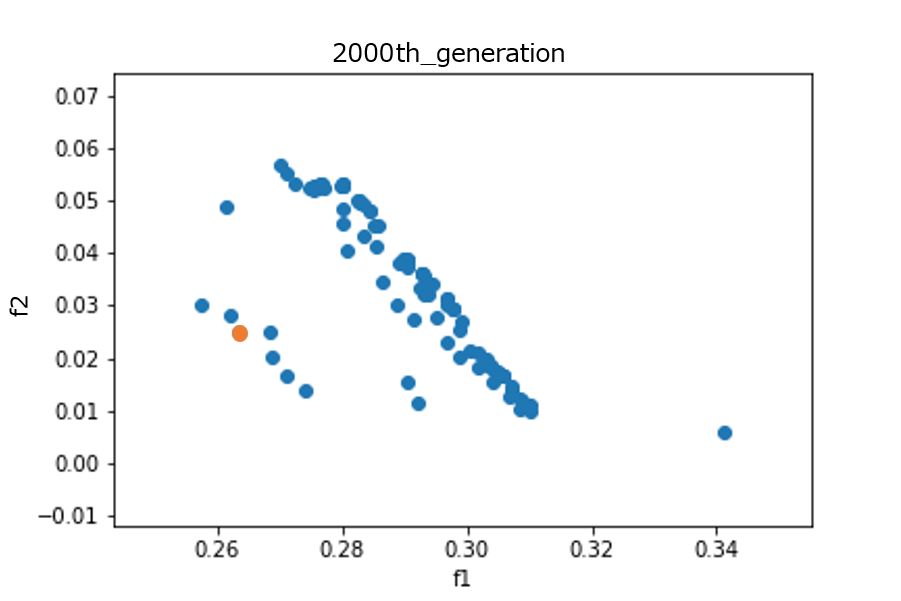} &
    \includegraphics[clip,width=35mm]{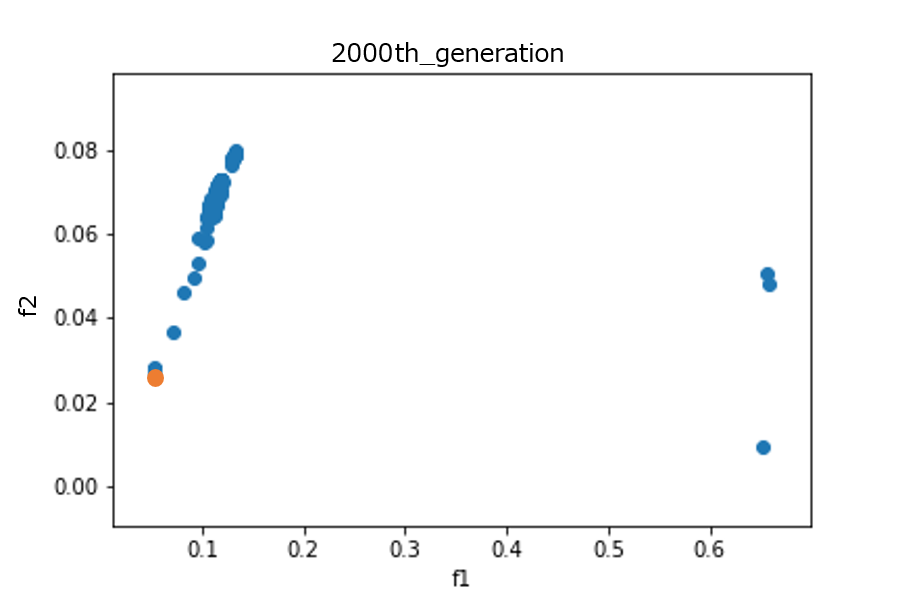} &
    \includegraphics[clip,width=35mm]{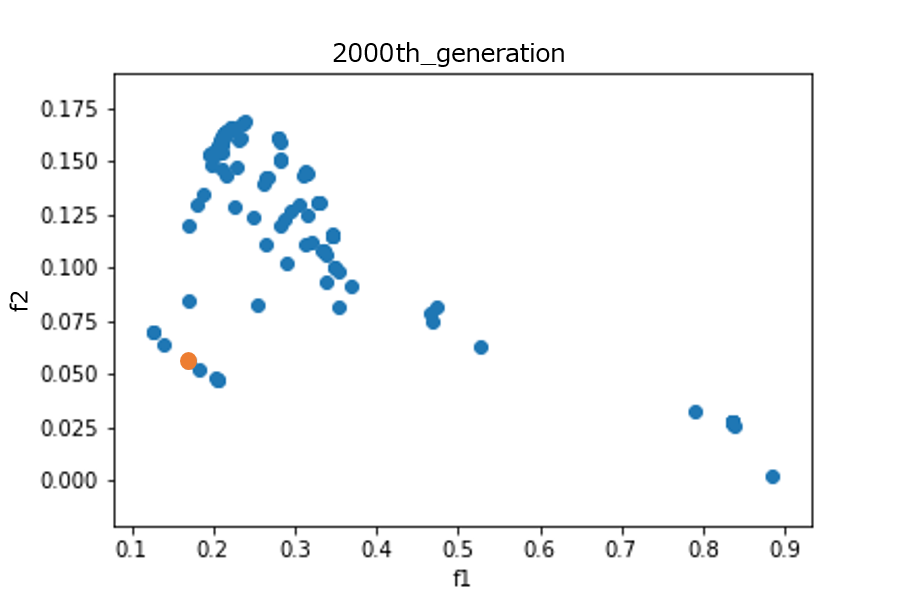} &
    \includegraphics[clip,width=35mm]{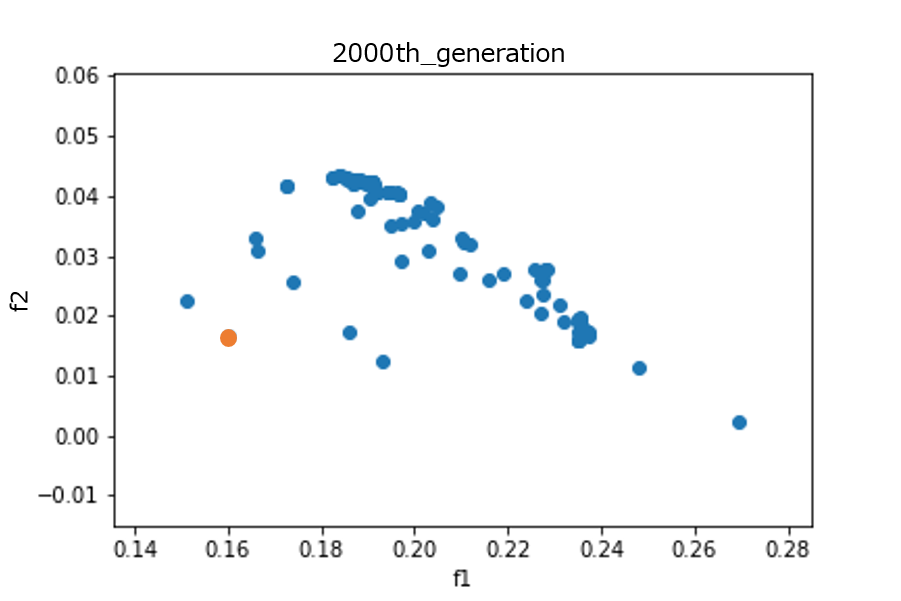} \\
    {\small (f) ``stop'' } &
    {\small (g) ``up'' } &
    {\small (h) ``yes'' } &
    {\small (i) ``left'' } &
    {\small (j) ``go'' } \\
    \end{tabular}
    \caption{Obtained non-dominated solutions.}
    \label{fig:select_individual}
~\\~\\  
%
    \centering
    \begin{tabular}[t]{@{}c@{}c@{}c@{}c@{}c@{}}
    \includegraphics[clip,width=35mm]{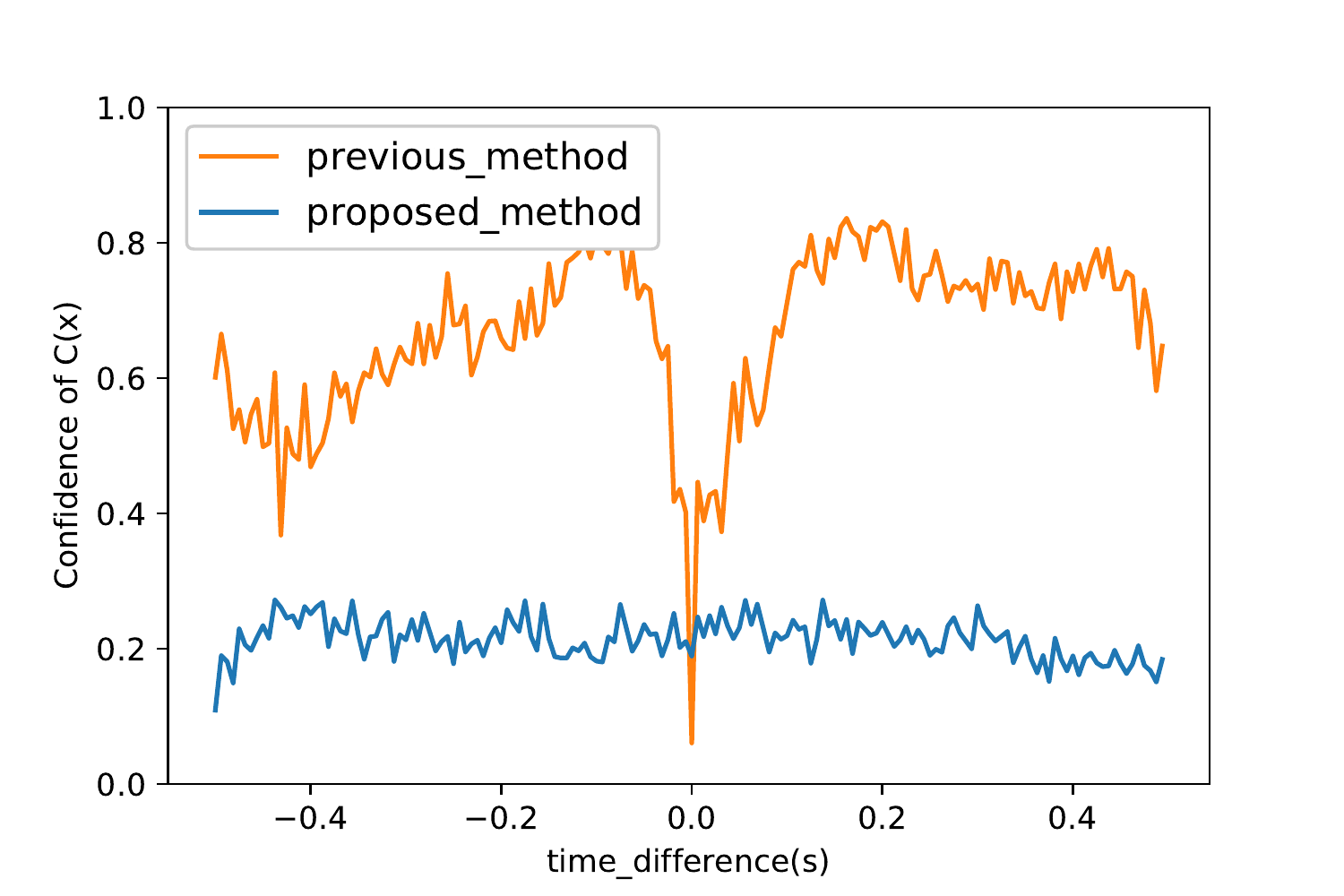} &
    \includegraphics[clip,width=35mm]{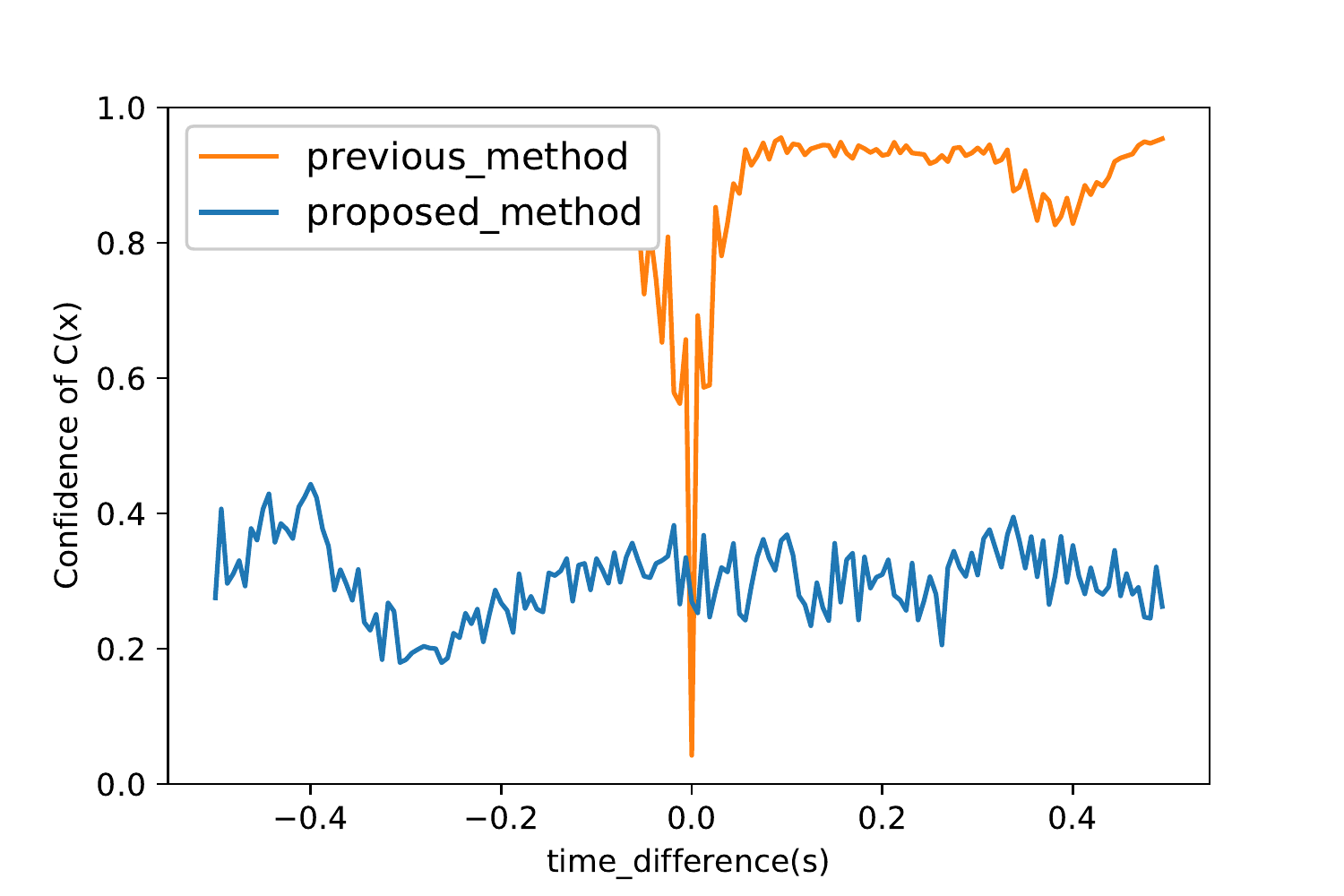} &
    \includegraphics[clip,width=35mm]{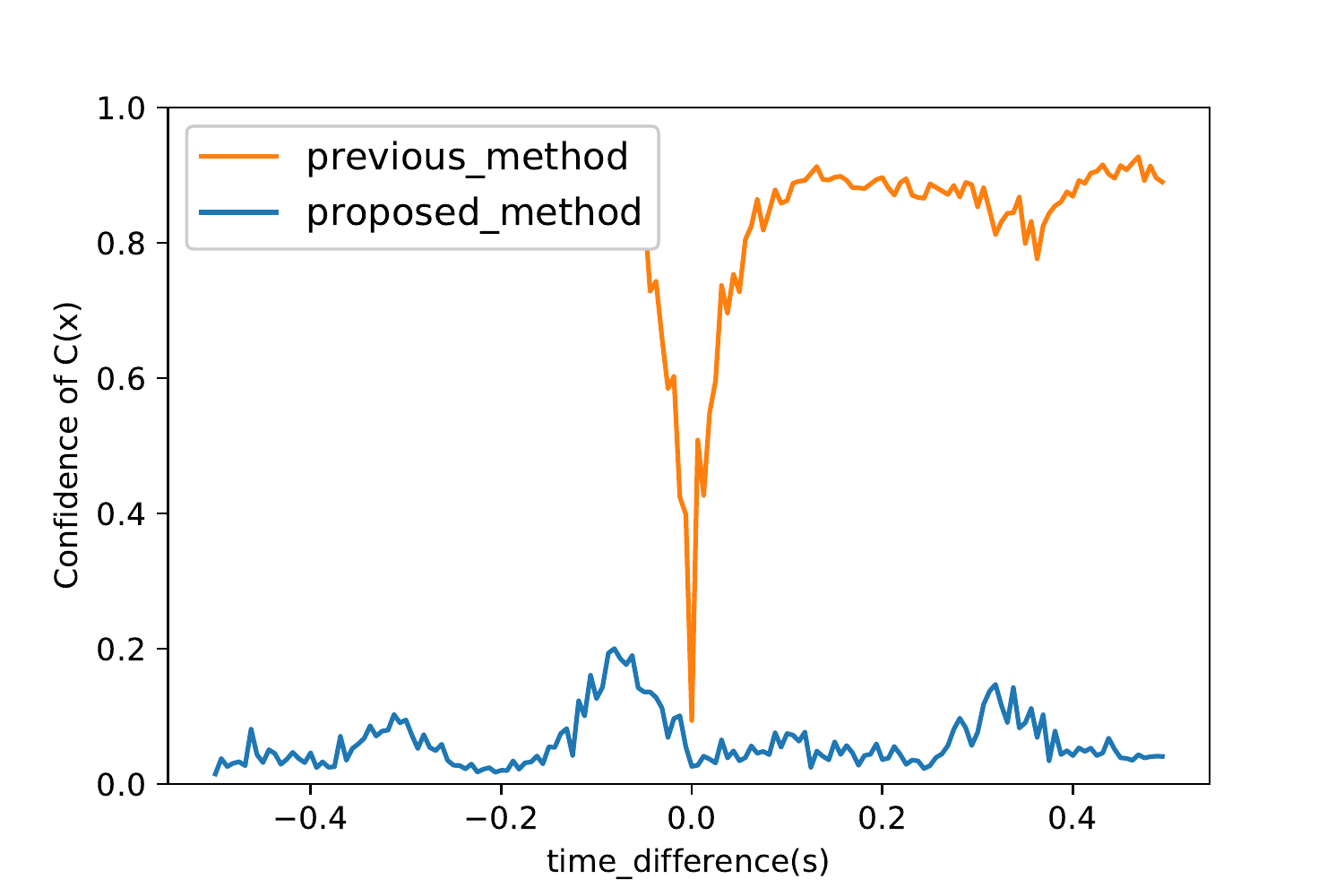} &
    \includegraphics[clip,width=35mm]{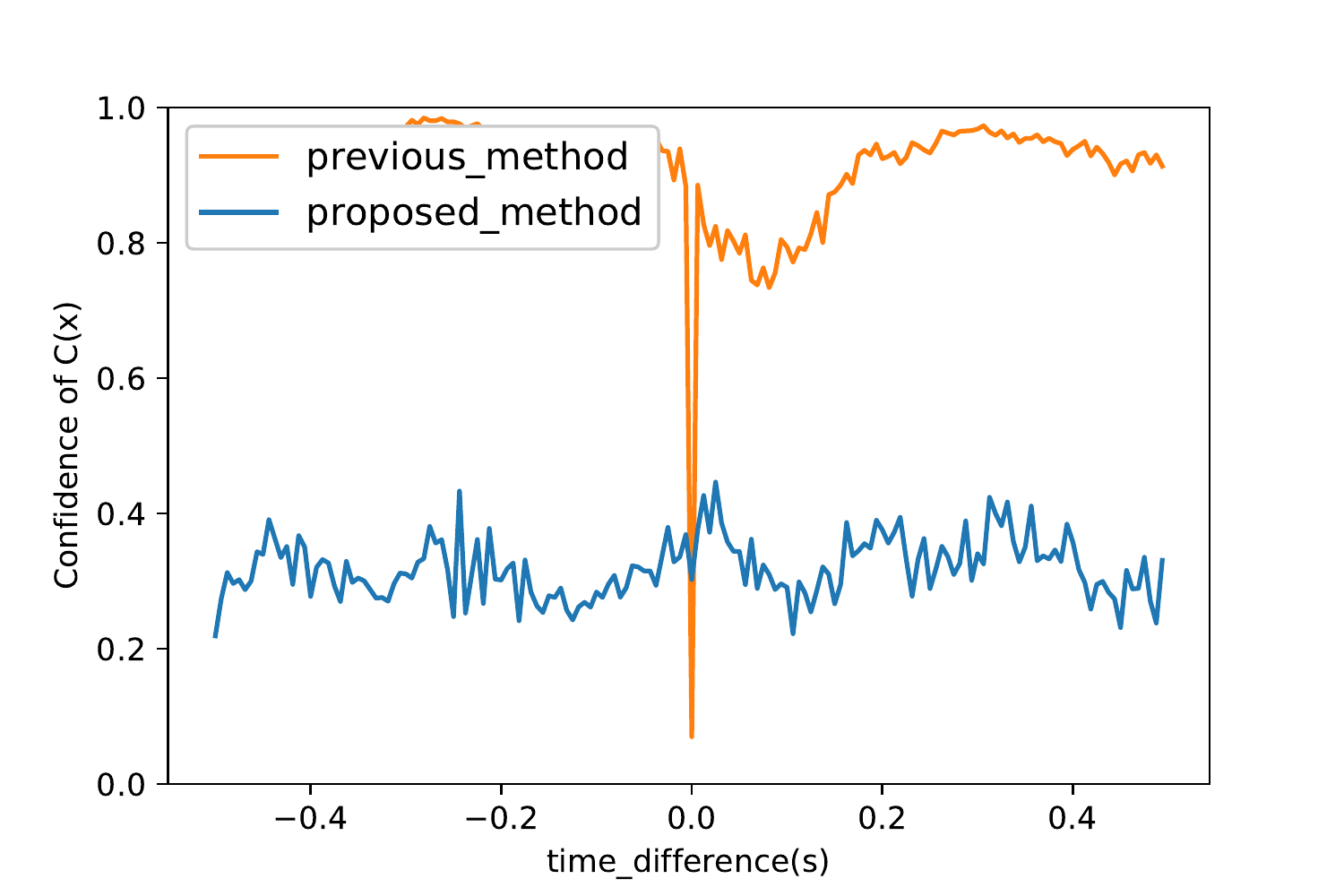} &
    \includegraphics[clip,width=35mm]{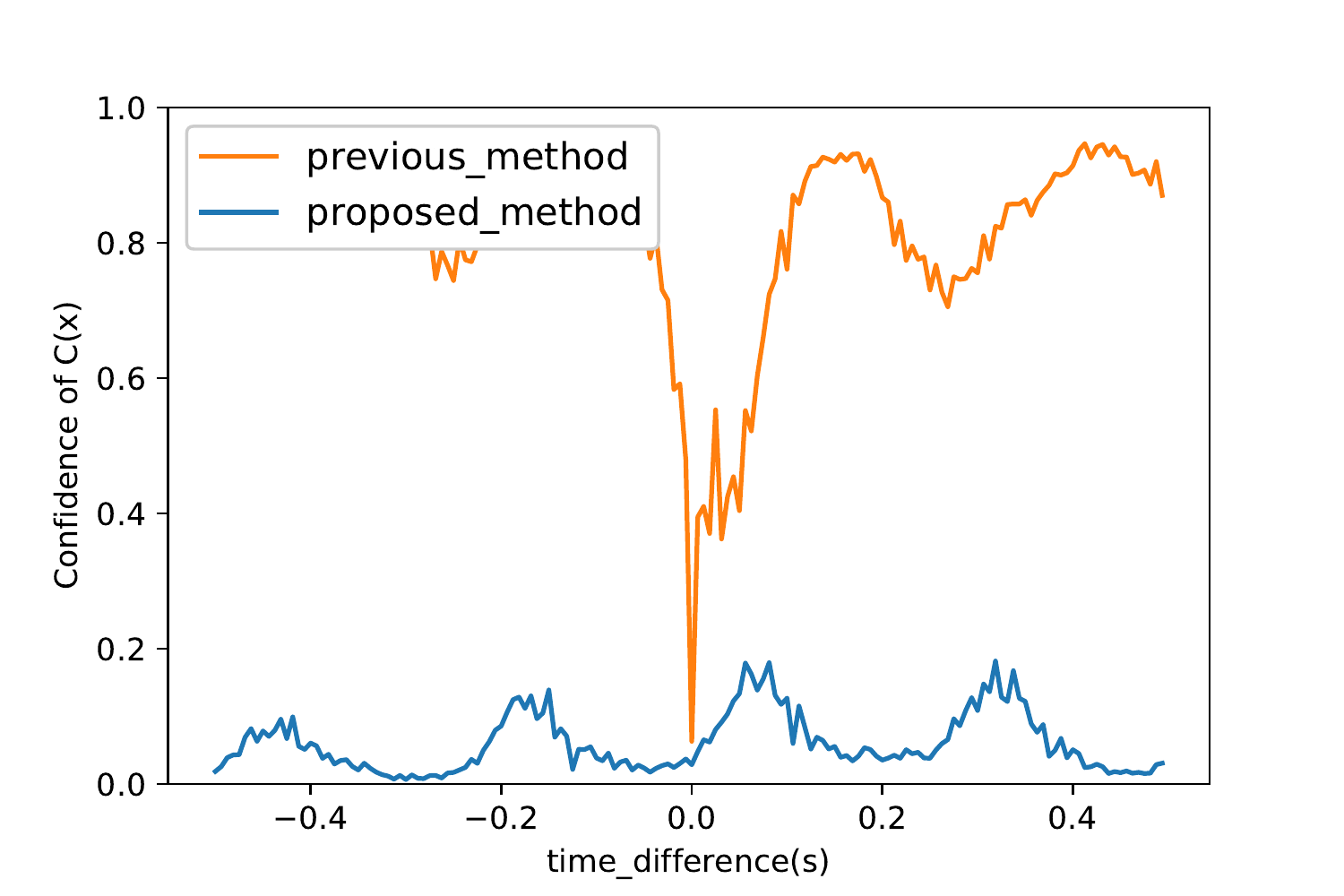} \\
    {\small (a) ``down'' } &
    {\small (b) ``no'' } &
    {\small (c) ``off'' } &
    {\small (d) ``on'' } &
    {\small (e) ``right'' } \\
    \includegraphics[clip,width=35mm]{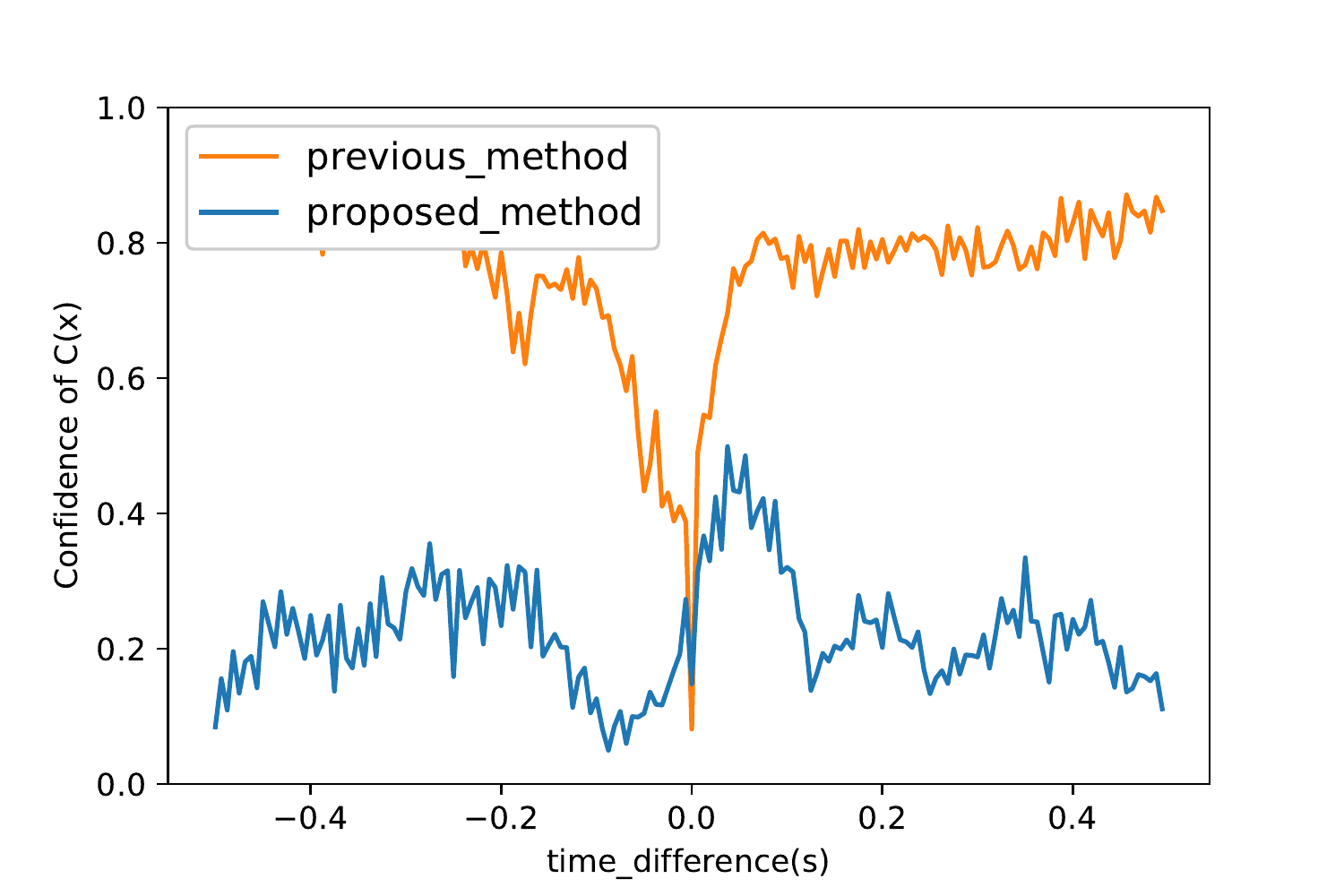} &
    \includegraphics[clip,width=35mm]{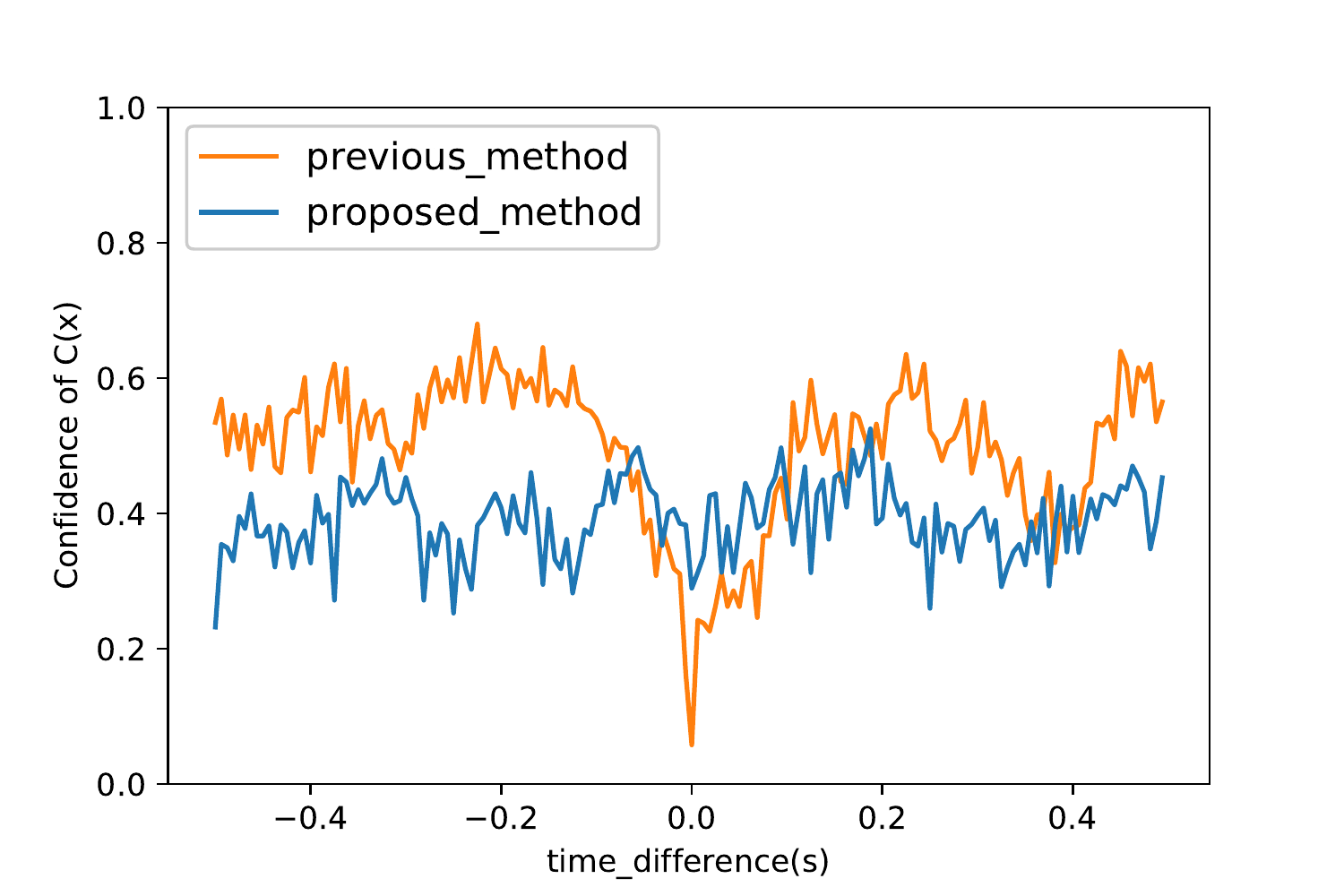} &
    \includegraphics[clip,width=35mm]{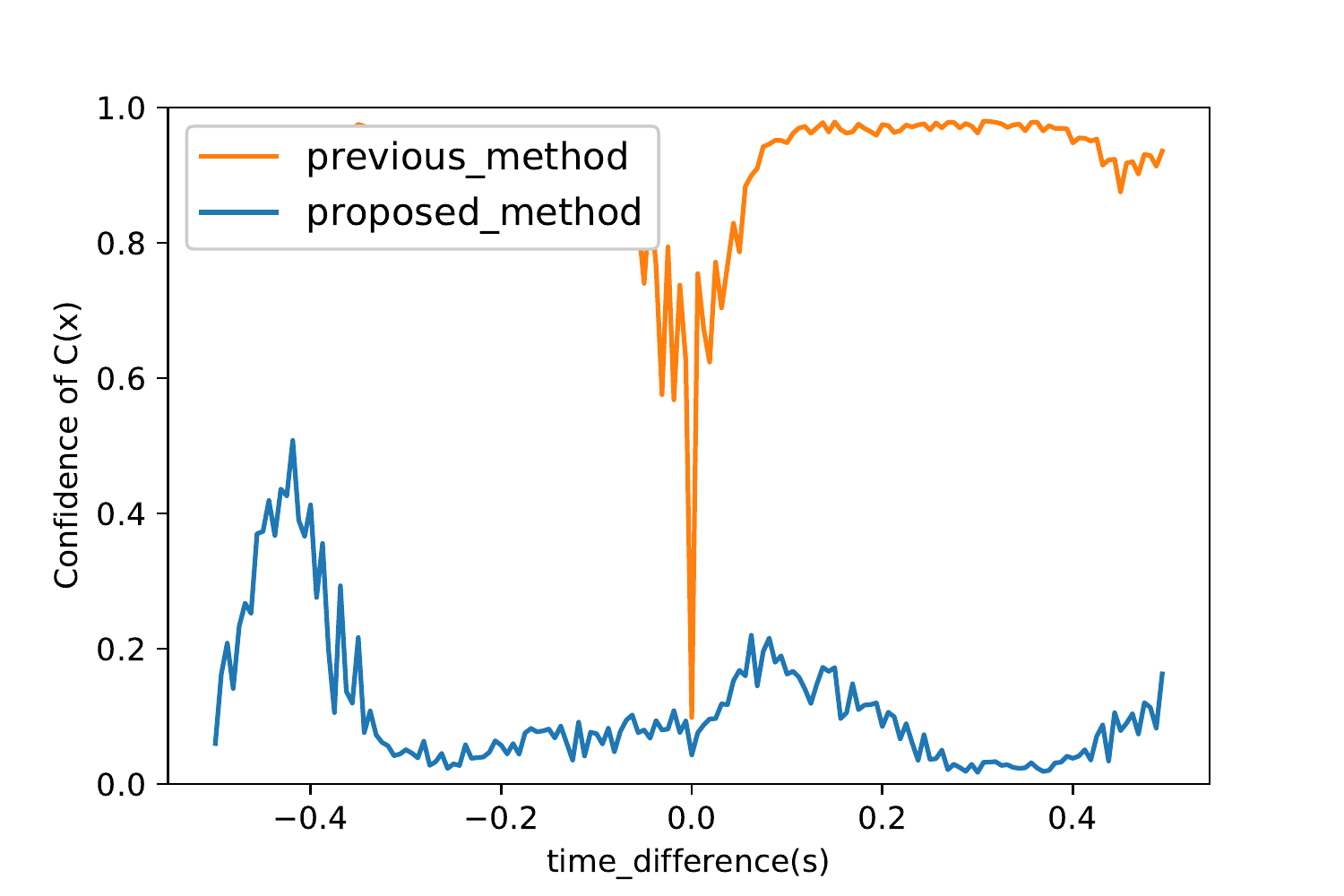} &
    \includegraphics[clip,width=35mm]{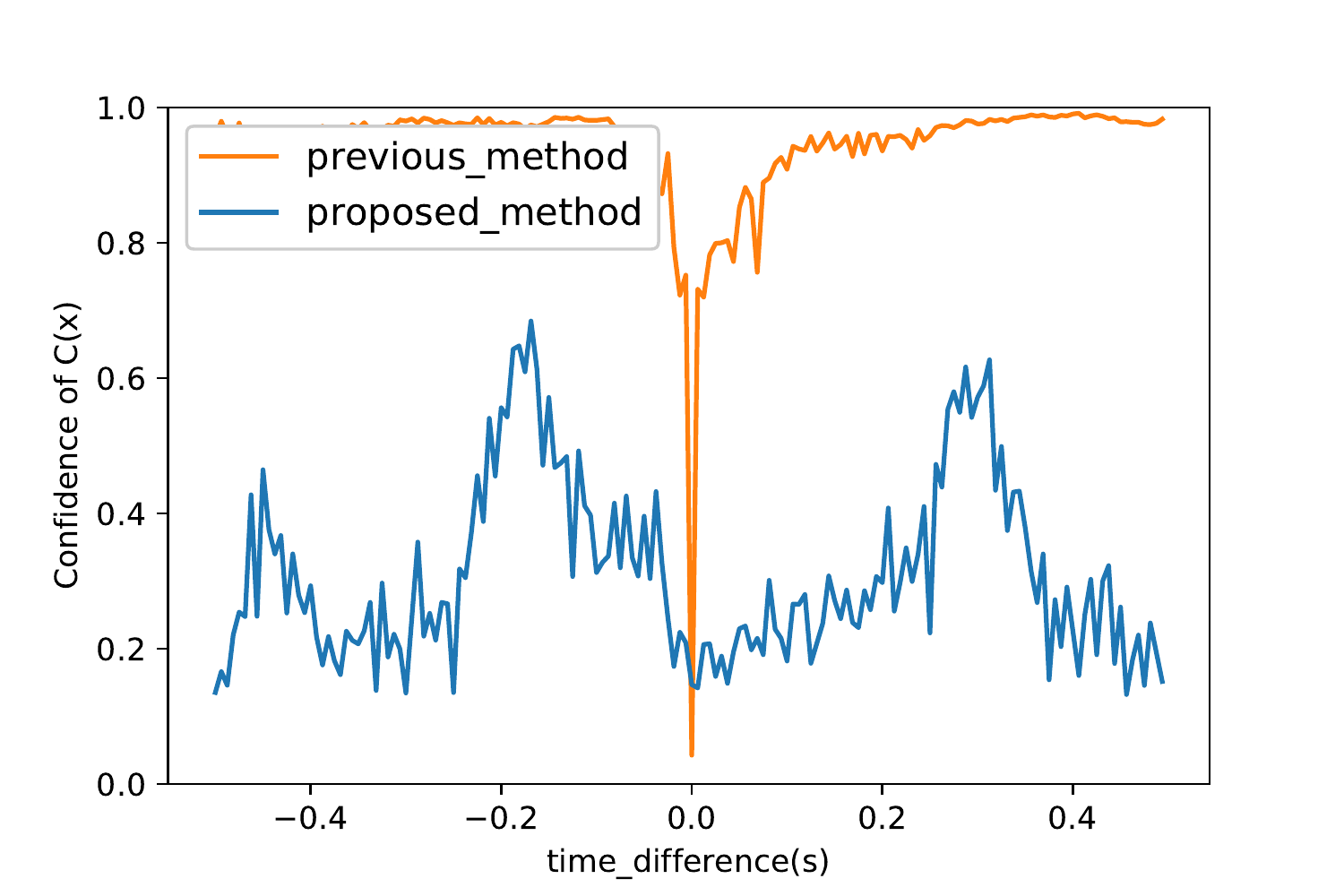} &
    \includegraphics[clip,width=35mm]{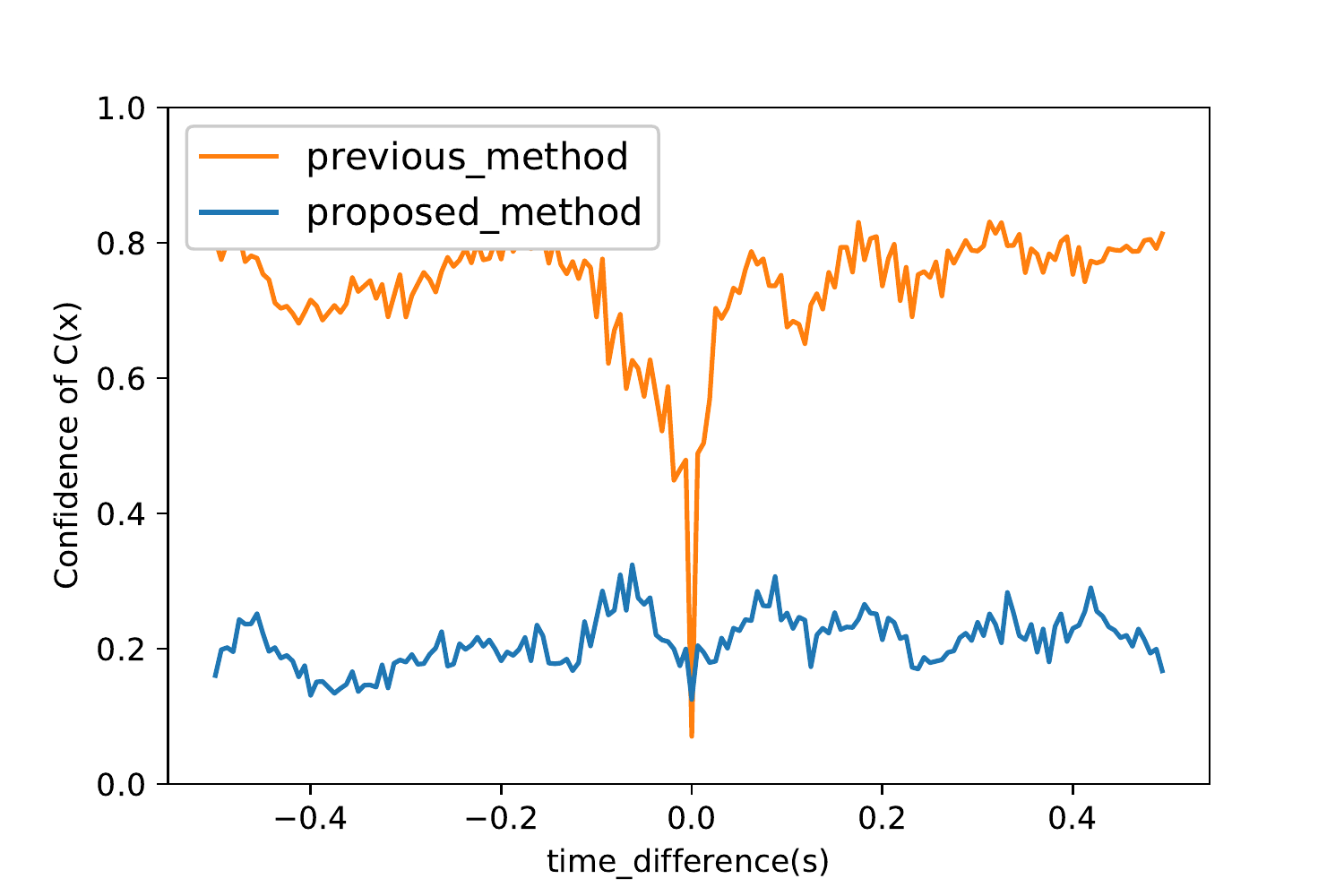} \\
    {\small (f) ``stop'' } &
    {\small (g) ``up'' } &
    {\small (h) ``yes'' } &
    {\small (i) ``left'' } &
    {\small (j) ``go'' } \\
    \end{tabular}
    \caption{Comparison on the robustness against timing lag between
    previous work~\cite{alzantot2018did} and the proposed method.}
    \label{fig:comparison_robustness}
  \end{figure*}

  \section{EVALUATION}
  
  \subsection{Experimental setup}
  \jtextd{
  提案する方式の有効性を検証するために、TensorFlowフレームワークで実装された音声コマンド分類モデル~\cite{sainath2015convolutional}を対象として敵対的事例の生成を試みた．
  音声コマンドデータセットは，65,000個のオーディオファイルを含むデータセットであり，"yes", "no", "up", "down", "left", "right", "on", "off", "stop", "go"のような単語の１秒のオーディファイルで構成されている．
  \cite{sainath2015convolutional}のモデルは，畳み込みニューラルネットワークに基づくモデルであり，音声コマンドデータセットで90\%の分類制度を実現しました．
  実験では提案手法と，遺伝的アルゴリズムを用いて敵対的音声事例を生成する手法~\cite{alzantot2018did}とを比較して行う．
  }
  
  In order to verify the effectiveness of the proposed method, we tried
  to generate adversarial examples for the speech command
  classification~\cite {sainath2015convolutional}.
  %
%
  As a target model, we adopt the trained model based on a
  convolutional neural network that achieved classification accuracy
  of 90\% in the speech command dataset~\cite
  {sainath2015convolutional}, which 
  contains 65,000
  samples of 10 classes such as  
  "off", "on", "right" and "stop."

%
  In the experiment, the proposed method is compared with the previous
  method for generating adversarial examples for ASR systems using
  genetic algorithm~\cite {alzantot2018did}.
  To design adjust-free perturbation, when evaluating solution
  candidates, the proposed method evaluates 9 perturbed sounds with
  lags within $T_{max} = 0.5$ [s] lags for each candidate.
  %
  Population size $N_{pop}$ and generation limit $N_G$ were set to 100 and 2,000,
  respectively.

  \subsection{Experimental results}
  \jtextd{
  
  提案手法の最適化過程における，解候補の目的関数空間における推移
  を図\ref{fig:transition}に示す．
  %
  最適化の序盤では摂動が大きく，また，対象モデルの誤認識を生じさせていないが，
  世代を重ねるごとに小さな摂動で安定的に対象モデルの誤認識を引き起こす敵
  対的事例が生成されていることがわかる．
  
  提案手法と従来手法とにより生成された敵対的事例を図
  \ref{fig:generated_example}に示す．
  提案手法は比較手法に比べ，時間軸方向に頑健になるために，振幅の大きさが全体的に一律な敵対的摂動が生成されていることがわかる．
  
  さらに，生成された敵対的事例の，時間ずれに対する頑健性を図
  \ref{fig:comparison_robustness}に示す．
  図\ref{fig:comparison_robustness}に示す各グラフの横軸は，攻撃対象とす
  る入力音声に対する敵対的摂動の時間のずれを示し，縦軸は入力音声の正しい
  クラスの信頼度を表す．
  %
  提案手法は比較手法に比べ，全クラスにおいて，時間軸方向の幅広い範囲で
  信頼度を低下させることに成功しており，
  頑健な敵対的摂動が生成できていることが分かる．
  
  }
  
  Figure~\ref{fig:transition} shows the transition of the solution
  candidate distribution in the objective function space during the
  optimization by the proposed method.
  In the early step of the optimization, perturbations were large and
  did not cause misrecognition of the target model; however, after 250
  generations, the confidence of the correct class to the input
  began to fall below 0.5 while their perturbation amount decreased.
%
  %
  Figure~\ref{fig:generated_example} shows the obtained adversarial
  examples for class ``down'',
  which reveals that the one generated by the proposed
  method includes the smoother pattern than that by the previous method.
  
  \jtextd{
  全クラスのf1-f3の目的関数空間を示している．
  横軸が～縦軸が～
  左下にいけばいくほどロバストであると言える．
  }
  
  Figure~\ref{fig:select_individual} shows the obtained non-dominated
  solution distributions projected to $f_1$-$f_2$ plane.
  %
  Solutions located at the bottom left of the graphs are more robust
  i.e., more stable in causing misrecognition.
%
  %
%
  A representative robust solution for each class is selected from
  the bottom-left part of the distribution and compared with ones
  generated by the previous method in terms of the robustness against
  time difference to the target speech as shown in
  Figure~\ref{fig:comparison_robustness}.
  The horizontal axis of each graph shows the time difference
  between adversarial perturbation and the input speech to be attacked,
  and the vertical axis shows the confidence of the correct class of the
  input.
  \jtextd{これらのグラフを書くために，Figure~\ref{fig:select_individual}の各クラスからオレンジ色の解，左下付近の解，サンプルを選択した．
  }
  %
%
  Although the confidence score of the adversarial examples generated
  by the previous method were sufficiently low when the lag was almost
  zero, the effect of the attack rapidly deteriorates when there is
  even a small time difference.
  On the other hand, the proposed method succeeded in
  reducing the confidence over the whole range in the tested time
  difference in almost all classes except ''left'' class where the
  generated examples failed to fool the classifier in some ranges such
  as from -0.25 to -0.15 [s] and 0.25 to 0.35 [s]. 

  Because the target sound length is 1.0 seconds, the generated
  adversarial examples that could lead the misclassification from -0.5
  to 0.5 [s] time lag can be recognized as adjust-free. 
  That is, for instance, by repeatedly playing the perturbation sound,
  the 0.7 [s] time difference can be regarded as -0.3 [s] lag.
  Therefore, the proposed method succeeded in generating the
  adjust-free adversarial examples in six over 10 classes when
  assuming the confidence threshold of 0.49 that was the average of
  the maximum confidence except the correct class.
%

  
  \subsection{Effect of Minimize standard deviation}
  \jtextd{
  （分類制度の信頼度や類似度だけを目的関数として最適化をしていくと，ある時間では信頼度が高いままということが懸念される．つまり～ということがおきる．そのため，提案手法では信頼度や類似度だけでなく，各評価点における分類制度の標準偏差も目的関数として加えている．）
  %
  提案手法では，正しいクラスの信頼度の期待値に加えて，標準偏差を目的関数に加えて最小化を行ってる．
  本節では標準偏差の最小化がもたらす影響を述べる．
  図～はクラス"down"において，最終世代のパレート解のうち，
  期待値f1が同等でかつ標準偏差f2が異なる2つの個体の時間ずれのグラフを表している．f2が大きい方の個体は，+0.2~+0.4秒すれることによって，f1の値が0.5を超えていることが分かる．しかし，f2が小さい個体は全ての時間を通して，f1が0.5を下回る滑らかなグラフとなっている．
  このことから，
  任意の量の
  時間ずれが生じた際に安定的に誤認識を引き起こす
  ajust free perturbationを作成するためには，信頼度の標準偏差を目的関数として含める
  ことが有効であることがわかる．
  \onote{【もっともよい方法は$f_2$（標準偏差）を外して2目的で行う方法と
      比較することなので，もしかすると査読者よりそのような追加実験の指示
      があるかもしれません．】}
  }
  
\begin{figure}[t]
    \centering
    \includegraphics[clip,width=50mm]{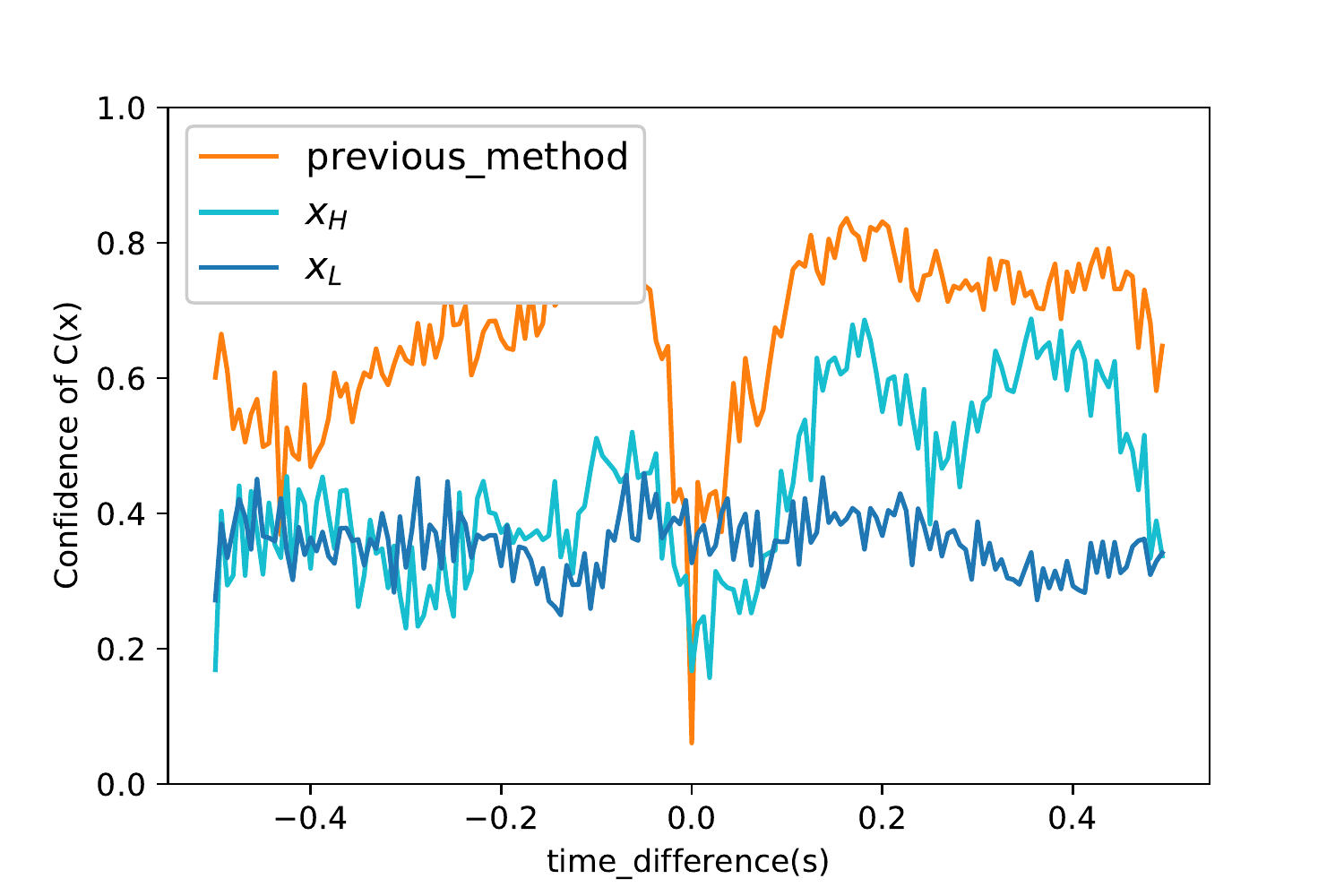} \\
    \caption {The effectiveness of employing standard deviation.}
    \label{fig:standard deviation}
  \end{figure}

  %
  Because the proposed method minimizes the standard deviation of the confidence
  of the correct class in addition to the expected value,
  this section describes the effects of minimizing the standard deviation.
  Figure~\ref{fig:standard deviation} shows the non-dominated solutions
  designed for an input speech belonging to class ``down.''
  The robustness against the time difference of the two individuals
  $\Vec{x}_L$ and $\Vec{x}_H$ that have the same expected value $f_1$ but
  different values of standard deviation $f_2$.
  %
  $\Vec{x}_H$ sometimes failed to make the classifier fool when
  there was a time lag of about $+0.2$ to $+0.4$ seconds, whereas
  $\Vec{x}_L$ successfully mislead the classifier throughout the time
  lag from $-0.5$ to $+0.5$, achieving  generating the adjust-free adversarial 
  perturbation.
  %
  %
  Therefore, it is effective to use the standard deviation of confidence
  as one of objective functions to create robust adversarial
  examples against the time lag.
  
  \ocut{
  \subsection{Discussion}
  \jtext{
  このセクションでは，将来の方向性について説明する．
  
  より大きなASRモデル，データセットに対する評価：より強力な最新のASRモデルでも敵対的事例の影響を受けるか、および対象とするデータセットが単一の単語のオーディオクリップではなく，会話のような文のオーディオの場合でも生成できるかどうかだ．
  
  現実世界での攻撃：本論文の評価では，攻撃者がオーディオクリップを分類モデルに直接入力すると仮定している．現実世界において，対象モデル（商用デバイス）に対し，スピーカーから敵対的音声事例を再生する場合でも攻撃が成功すれば，より現実的かつ強力な攻撃といえる．
  }
  }

  \section{CONCLUSION}
  
  \jtextd{
  本論文では，ASRシステムに対して進化型多目的最適化を用いた頑健な敵対的
  事例の生成手法を提案した．提案手法は，対象モデルの内部情報を必要としな
  いブラックボックス手法であり，トレードオフの関係にある複数の目的関数を
  同時に最適化することで，
  時間軸方向に頑健な敵対的事例が生成で
  きる．
  今後，問題の次元削減や他モデルへの転用
  について検討する．
  }
  
  This paper proposes a method to generate robust adversarial examples
  using evolutionary multi-objective optimization for ASR systems.
  The proposed EMO-based black-box attack method 
  generates robust adversarial examples against the
  time difference to target speech by minimizing
  the standard deviation of the confidence score in addition to the
  expectation of it.
  Experimental results showed that the proposed method could generate
  adjust-free adversarial examples, which are sufficiently robust
  against the time difference so that attackers do not need to
  determine the timing of playing the adversarial noise sound. 
  In future, we plan to design a problem dimension reduction method and
  verify the effectiveness of our method for other ASR models.

  \begin{acknowledgements}
    This study was partially supported by the Kayamori Foundation of
    Informational Science Advancement.
    
  \end{acknowledgements}
  
  \bibliographystyle{spphys}       
  \bibliography{ref}   
  
  \end{document}